\documentclass[letterpaper]{article}
\usepackage{hyperref}%

\def\Psih{\Psi^{\mathrm{hor}}}

\def\tPsi{{\tilde \Psi}}

\def\cF{{\mathcal F}}
\def\cU{{\mathcal U}}
\def\cW{{\mathcal W}}

\begin{document}


\centerline{\LARGE Quantizing Gravitational Collapse}
\vskip 3mm
\centerline{Cenalo Vaz}
\vskip 3mm
\centerline{\it University of Cincinnati, Cincinnati, OH 45221-0011}
\centerline{\it and}
\centerline{\it Universidade do Algarve, Faro, Portugal.}

\section{Introduction}

It is generally believed that the following is true: the classical gravitational field of a
sufficiently massive star will overcome its neutron degeneracy pressure and, barring quantum
gravitational effects in the final stages of collapse, the star will form a singularity of
space-time. The initial conditions determine the precise nature of the singularity of the
collapse: either the singularity will be covered, in which case a black hole is formed, or
it will be naked, {\it i.e.,} not covered by a horizon. Naked singularities lead to violations
of causality, which encouraged Penrose to propose the Cosmic Censorship Hypothesis
(CCH)\cite{rp}. The CCH simply states that physically reasonable initial data cannot produce
naked singularities. Nevertheless, to date there is little evidence that it is true in general
relativity, even though most of our current emphasis on black hole physics is based on the
validity of this hypothesis.

The LeMa\^\i tre-Tolman-Bondi (LTB) collapse of spherical, inhomogeneous dust\cite{ltb} is
(arguably) the simplest model that allows for the formation both of black holes and of naked
singularities. It is defined by specifying two functions, {\it viz.,} a mass function, $F(\rho)$,
and an energy function, $f(\rho)$, where $\rho$ is a shell label coordinate. The former
represents the weighted mass contained within the matter shell labeled by $\rho$, and the latter
is related to the velocity profile within the collapsing cloud at the initial time. The
configuration space for all LTB models consists of the dust proper time, $\tau$, and the
area radius, $R$. The classical collapse may end in a black hole or in a naked singularity depending on the
behavior of $F(\rho)$ and $f(\rho)$ near the center\cite{psj1}. When it ends in a black
hole, a significant fraction of the star is expected to evaporate via Hawking radiation
\cite{haw} during the semi-classical phase. However, the back-reaction of the space-time
prevents us from making definitive predictions at the final stages. A different result is
obtained when the collapse is toward a naked singularity\cite{cv1}. It is quantum
mechanically unstable (the radiated power behaves as $P \sim (U_0-U)^{-\alpha}$ for some
-- model dependent -- value of $\alpha > 0$)  but, during the validity of the semi-classical
approximation (curvatures should be less than Planck scale), the collapsing cloud emits only
about one Planck unit of energy\cite{cv2}. Because the back-reaction does not become important
so long as gravity can be treated classically, the future evolution of the star is governed
{\it exclusively} by quantum gravity and it is impossible to say, from the semi-classical
approximation, whether the star radiates away its energy on a short time scale or settles down
into a black hole state. Quantum gravitational effects are therefore expected to modify the
very nature of the classical singularities that result from the collapse of a relativistic star,
serving even as the Cosmic Censor. However, the true end state of the collapse can only be
fully understood by an application of a quantum theory of gravity.

\section{Quantization}

The canonical dynamics of the collapsing cloud is described by embedding the spherically
symmetric ADM\cite{adm} 4-metric in the LTB space-time, and by casting the action for the
Einstein-dust system in canonical form. Performing a version of the canonical transformation
developed by Kucha\v r\cite{kk1}, one may use the chart consisting of $(\tau, R,F,P_\tau,
P_R,P_F)$ whence the time evolution of the wave-functional, $\Psi[\tau,R,F]$, is determined
by the functional Schroedinger equation,
\begin{equation}
i\frac{\delta \Psi}{\delta \tau} = {\hat h}\Psi = \sqrt{\pm c^2 \frac{\delta^2}{\delta R_*^2} \pm
m_p^2 c^4 \frac{{F'}^2}{4|\cF|}}~ \Psi[\tau,R, F],
\label{dyn}
\end{equation}
the upper sign within the square-root referring to the region outside the horizon and the lower sign
to the region inside. Invariance under spatial diffeomorphisms is implemented by the momentum
constraint,
\begin{equation}
\left[ \tau ^{\prime }\frac \delta {\delta \tau }+R_{*}^{\prime }\frac \delta {\delta R_{*}}+F^{\prime }
\frac \delta {\delta F}\right] \Psi [\tau,R,F]=0\
\label{mc}
\end{equation}
and an inner product on the Hilbert space of wave-functionals may be defined in a natural way, by
exploiting the fact that the DeWitt super-metric\cite{dw} is manifestly flat in the configuration space
$(\tau,R_*)$, as the functional integral\cite{cv3}
\begin{equation}
\label{ip}\left\langle \Psi _{1}\vert\Psi _2\right\rangle =\int_{R_{*}(0)}^\infty \mathcal{D}R_{*}
\Psi^\dagger_1\Psi _2\ .
\end{equation}
This inner product is defined on  $\tau =$ constant hypersurfaces and the set-up in equations
(\ref{dyn})--(\ref{ip}) implies a specific choice, albeit a natural one, of operator ordering. For
a solution to the quantum constraints, we note that momentum constraint  is obeyed by {\it any}
functional that is a spatial scalar. In particular the functional
\begin{equation}
\Psi = A \exp \left[-\frac{ic}{2l_p^2}\int_0^\infty dr F' \left[\tau + \cU(R,F)
\right] \right],
\label{ansatz}
\end{equation}
where $A=A(F)$, is a spatial scalar if $\cU(R,F)$ has no explicit dependence on $r$ and describes
a stationary state of proper energy $F'/2$. Throughout we will use this as a solution ansatz\cite{cv4}.

\section{The ``eternal" black hole: Mass Spectrum and Entropy}

A mass function describing an eternal black hole of mass $M$ may be taken to be that of a
spherical shell $F(r) = 2M \theta (r)$, where $\theta(x)$ is the Heaviside function.
Equation (\ref{ansatz}) then tells us that the problem of a single shell is essentially
quantum {\it mechanical},
\begin{equation}
\tPsi[\tau,R,F] = \exp\left[M \cW_0(\tau,R,M)\right],
\label{sshell}
\end{equation}
where $\tau = \tau(0), R = R(0)$ and $F(0)=2M$ represent, respectively, the proper time, the radial
coordinate and the total mass of the single shell. In the WKB approximation, the stationary states
of the black hole are easily described: in the interior they are a superposition of ingoing and
outgoing waves,\cite{cv4}
\begin{equation}
\tPsi[\tau,R_*] = A_\pm e^{-iM(\tau\pm R_*)},~~~~~ -\pi M < R_* < \pi M
\label{in}
\end{equation}
and in the exterior, they are exponentially decaying
\begin{eqnarray}
\tPsi[\tau,R_*] &=& B e^{-iM(\tau -i R_*)},~~~~~ R_* > \pi M\cr
&=& C e^{-iM(\tau + iR_*)},~~~~~ R_* < -\pi M,
\label{out}
\end{eqnarray}
which is in keeping with the fact that there are no wave solutions in vacuum spherical
gravity. Matching the wave function and its derivative at the horizon, one finds that the
energy (mass) squared of the black hole is quantized in half integer units,
\begin{equation}
M^2 = \left(n+\frac{1}{2}\right) M_p^2,~~ \forall~~ n \in {\bf N} \cup \{0\}
\end{equation}
where $M_p$ is the Planck mass. This is just the mass spectrum originally proposed by
Bekenstein\cite{bek1}.

The above considerations may be extended to the case of many shells. A simple generalization
of the mass function above, which describes many shells, is $F(r) = 2\sum_{j=1}^N \mu_j
\theta(r-r_j)$ and a straightforward application of the boundary conditions appropriate to
each shell gives the following quantization condition for the states of shell $k$,\cite{cv5}
\begin{equation}
\mu_k M_k = \left(n_k + \frac{1}{2} \right) M_p^2.
\label{shellcond}
\end{equation}
where $M_k$ represents the total mass contained within shell $k$. These conditions, if
applied recursively, show that the mass of shell $k$ is determined by $k$ quantum numbers.
Thus the total mass depends on $N$ quantum numbers for a black hole
formed out of $N$ quantum shells. The appearance of $N$ quantum numbers means a quantum
black hole is not simply described by its total mass: such a description ignores the manner
in which the mass is distributed among the shells. The entropy counts the number of
distributions for a given total mass. For an {\it eternal} black hole, $M_k$ in (\ref{shellcond})
should be replaced by $M$, the mass of the hole. The total mass (squared) of the hole
continues quantized as before and the problem of counting the number of distributions is
precisely the problem of asking for the number of ways in which $N$ integers may be added
to give another integer\cite{cv5}. This result depends on the number of shells that have collapsed
to form the black hole, which we do not know but which can be independently determined by
maximizing the entropy with respect to $N$. When both $N$ and $M/M_p$ are large, one readily
finds, to leading order,
\begin{equation}
S \approx 0.962 \times \frac{\mathcal A}{4}
\end{equation}
in units of Planck area, which agrees well with the Bekenstein-Hawking value.

\section{Hawking Radiation}

When applied to a matter distribution that is appropriate to a massive black hole surrounded
by dust whose total energy is small compared with the mass of the black hole, the WKB approximation
describes Hawking radiation. For this purpose,
let us assume that the mass function $F(r)$ is of the form\cite{cv6} $F(r)=2M\theta (r)+f(r)$,
where $\theta (r)$ is the Heaviside step-function, and $f(r)$ is differentiable, representing
a dust distribution with $f(r)/2M\ll 1.$ It can be interpreted as the presence
of a Schwarzschild black hole of mass $M$ at the origin, and $f(r)$ is a dust matter
perturbation on the black hole, which can be related to Hawking radiation. In a sense, it
plays the role of the quantum field used in standard derivations of Hawking radiation, except
that now it is quantized along with the gravitational field.

Inserting this mass function in the wave-functional (\ref{ansatz}) we have $\Psi =\Psi_{\rm bh}
\times \Psi_f$ where $\Psi_{\rm bh}$ represents the black hole and
\begin{equation}
\nonumber
\Psi_f \sim \exp \left[\frac 12\int_0^\infty drf^{\prime }{\mathcal W}^f(\tau(r),
R(r),M)\right]\ .
\end{equation}
The solution for ${\mathcal W}^{f}$ outside the apparent horizon and written in terms of
Schwarzschild time is given by
\begin{equation}
{\mathcal W}^{f}_{\rm out} = -i\left[ T + 4\sqrt{2M}\left( \sqrt{R}-\frac{\sqrt{2M}}2\ln
\left[ \frac{\sqrt{R}+ \sqrt{2M}}{\sqrt{R}-\sqrt{2M}}\right] \right) \right]\ .
\label{asp}
\end{equation}
The projection of our solution (\ref{asp}) on the negative
frequency modes of an outgoing basis on ${\mathcal I}^{+}$ represents the negative frequency
modes present in the solution. So we compute $|\langle \Psi _\omega ^{\dagger}\vert\Psih_f\rangle |^2$
because these are the analogs of the Bogoliubov coefficients near the apparent horizon,
$|\beta (f,\omega )|^2.$
We find\cite{cv6}
\begin{equation}
\label{haw}|\beta (f,\omega )|^2=\prod_r\frac{2\pi M}{\Delta f}\left[
\frac 1{e^{8\pi M\Delta f}-1}\right].
\end{equation}
This is interpreted as the eternal black hole being in equilibrium with a thermal
bath at the Hawking temperature $(8\pi M)^{-1}.$ Our derivation provides a functional
Schr\"odinger picture for dust Hawking radiation, consistent with the WKB
wave-functional which solves the Wheeler-DeWitt equation.

\section{Beyond the WKB approximation}

Because of Hawking radiation, we do not expect to have a sharp boundary between the
collapsing matter and an external vacuum (as is generally considered in the classical
studies) owing to the evaporation process. Thus $F(r)$ will increase with increasing
$r\in [0,\infty)$ and we could equivalently label shells by $F(r)$ instead of $r$.
The Hamiltonian constraint is ill-defined and requires regularization,
which we perform on a lattice (for details on the lattice construction, see\cite{cv9}).
If $j$ represent lattice points, the wavefunctional can be expressed as the product
\begin{equation}
\Psi[\tau,R] = \prod_j \Psi_j(\tau_j,R_j) = \prod_j \exp\left[-\omega_j\left[
\tau_j + \cU_j(R_j,F_j)\right]\right].
\label{pprod}
\end{equation}
where $\varepsilon_j=\hbar \omega_j$ is the energy of shell $j$ and the Wheeler-DeWitt
equation is to be solved independently at each label. Each lattice point physically
represents a shell of matter. Thus we get an infinite set of formally identical equations,
one for each shell. After carrying out the time derivative , defining the dimensionless
variables $z_j=R_j/F_j$ and $\gamma_j = F_j\omega_j/c$, denoting $y_j\equiv \tPsi_j$
(the time independent shell wave function, and suppressing the subscript $j$ throughout,
the equation for each shell takes the form
\begin{equation}
z(z-1)^2 \frac{d^2 y}{dz^2} + \frac{(z-1)}{2}\frac{dy}{dz} +
\gamma^2 z^2 y= 0
\label{zeqn1}
\end{equation}
both in the exterior ($z>1$) and in the interior ($z<1$). Every shell's evolution is
therefore completely determined by the total mass contained within it.

Solutions to [\ref{zeqn1}] can be obtained as a power series about the non-essential
singularities at the origin ($z=0$) and the horizon ($z=1$). It is possible to show\cite{cv9}
that the Hawking spectrum itself arises from retaining only the lowest order terms and
dropping all terms of ${\mathcal O}(\hbar)$. When we include the
first order correction (in ${\mathcal O}(\hbar)$) we find,
\begin{equation}
|\beta(\omega,\omega')|^2 \approx 2\pi^2 F^2\sqrt{\frac{2c}{F\omega'}} \frac{kT_H}{\varepsilon} \frac{1}
{e^{\frac{\varepsilon}{kT_H}}-1}\left[1 - \frac{1}{2}\ln \left(\frac{\pi kT_H}{\varepsilon}
\right)\right],
\label{ntherm}
\end{equation}
which completes the first correction to the WKB approximation. The correction term cannot
be accounted for by modifying the Hawking temperature and one must conclude that it renders the
Hawking radiation non-thermal.

\section{Conclusions}

There are two points worth noting: (a) equation (\ref{ntherm}) does not represent a correction
to the Hawking temperature (equivalently, the Bekenstein-Hawking entropy). It shows instead
that the spectrum is {\it non-thermal}, and (b) this non-thermal spectrum {\it is associated
with the horizon} because we have considered only the near horizon wave-functional in our
calculation. It is not, therefore, related to the non-thermal spectrum observed at large
distances from the horizon, e.g., at spatial infinity, due to grey body factors arising from
the scattering of the radiation against the geometry.

Within the context and limitations of the canonical theory, our result indicates that unitarity
{\it is} preserved in quantum gravitational evolution\cite{hk}. Many open questions, such
as what meaning is to be assigned to the (logarithm of) the number of states, or the statistical
entropy, of the black hole in the absence of a thermal spectrum and what is the final fate of
black holes and naked singularities, remain to be examined.

\section{Acknowledgements}

Much of the work reported here was done in collaboration with Louis Witten and T.P. Singh. I
am very grateful for our close collaboration and many discussions. I also acknowledge the financial
support from the {\it Funda\c {c}\~ao para a Ci\^encia e a Tecnologia} (FCT) under contract
number POCTI/32694/FIS/2000 ({\it III Quadro Comunit\'ario de Apoio}).


\begin{thebibliography}{99}
\bibitem{rp}Roger Penrose, Riv. Nuovo Cimento {\bf 1} (1969) 252; {\it ibid.} in
{\it General Relativity, An Einstein Centenary Survey}, ed. S. W. Hawking and W. Israel,
Cambridge Univ. Press, Cambridge, England, (1979) 581.
\bibitem{ltb}G. LeMa\^\i tre, Ann. Soc. Sci. Bruxelles I, {\bf A53} (1933) 51; R.C.
Tolman (1934) Proc. Nat. Acad. Sci. USA {\bf 20} 169; H. Bondi (1947) Mon. Not. Astron. Soc.
{\bf 107} 410.
\bibitem{psj1}see, for example, P. S. Joshi, {\it Global Aspects in Gravitation and
Cosmology}, Clarendon Press, Oxford, (1993).
\bibitem{haw}S.W. Hawking, Comm. Math. Phys. {\bf 43} (1975) 199; S.W. Hawking, Phys. Rev.
{\bf D14} (1976) 2460; W.G. Unruh, Phys. Rev. {\bf D14} (1976) 870.
\bibitem{cv1}Cenalo Vaz and Louis Witten, Nucl. Phys. {\bf B487} (1997) 409; Cenalo Vaz
and Louis Witten, Phys. Lett. {\bf B442} (1998) 90; Cenalo Vaz and T.P. Singh, Phys. Rev.
{\bf D61} (2000) 124005.
\bibitem{cv2}T. Harada, H. Iguchi, K. Nakao, T.P. Singh, T. Tanaka and Cenalo Vaz, Phys. Rev.
{\bf D64} (2001) 041501.
\bibitem{adm}R. Arnowitt, S. Deser and C.W. Misner in {\it Gravitation: An Introduction to
Current Research}, ed. Louis Witten, Wiley, New York (1962).
\bibitem{kk1}K.V. Kucha\v r, Phys. Rev. {\bf D50} (1994) 3961.
\bibitem{dw}B.S. DeWitt, Phys. Rev. {\bf 160} (1967) 1113.
\bibitem{cv3}Cenalo Vaz, L. Witten and T.P. Singh, Phys. Rev. {\bf D63} (2001) 104020.
\bibitem{cv4}Cenalo Vaz and Louis Witten, Phys. Rev. {\bf D60} (1999) 024009.
\bibitem{bek1}J.D. Bekenstein, Lett. Nuovo Cimento {\bf 11} (1974) 467; {\it ibid}, Phys. Rev.
Lett. {\bf 70} (1993) 3680; {\it ibid}, Phys. Lett. {\bf B360} (1995) 7.
\bibitem{cv5}Cenalo Vaz and Louis Witten, Phys. Rev. {\bf D63} (2001) 024008.
\bibitem{cv6}Cenalo Vaz, Claus Kiefer, Louis Witten and T.P. Singh, Phys. Rev. {\bf D67}
(2003) 024014.
\bibitem{cv9}Cenalo Vaz, L. Witten and T.P. Singh, ``Exact Quantum State of Collapse and
Black Hole Evaporation'', \href{http://arXiv.org/abs/gr-qc/0306045}{[arXiv:gr-qc/0306045]}
\bibitem{hk}P. H\'aj\'\i \v cek, C. Kiefer, Nucl.Phys. B603 (2001) 531-554; P. H\'aj\'\i \v cek,
Nucl. Phys. {\bf B603} (2001) 555-577.


\end{thebibliography}
\end{document}